\documentclass[twocolumn]{aastex63}
\usepackage{newunicodechar}
\usepackage{amsmath}
\usepackage{color}

\DeclareUnicodeCharacter {2212} {-}
\DeclareUnicodeCharacter {02BC} {-}

\accepted{\today}
\submitjournal{ApJL}

\begin{document}

\title{ALMA Survey of Orion Planck Galactic Cold Clumps (ALMASOP): A forming quadruple system with continuum `ribbons' and intricate outflows}

\author[0000-0003-4506-3171]{Qiu-yi Luo}\thanks{E-mail:lqy@shao.ac.cn}
\affiliation{Shanghai Astronomical Observatory, Chinese Academy of Sciences, 80 Nandan Road, Shanghai 200030, People’s Republic of China}
\affiliation{School of Astronomy and Space Sciences, University of Chinese Academy of Sciences, No. 19A Yuquan Road, Beijing 100049, People’s Republic of China}

\author[0000-0002-5286-2564]{Tie Liu}\thanks{E-mail:liutie@shao.ac.cn}
\affiliation{Shanghai Astronomical Observatory, Chinese Academy of Sciences, 80 Nandan Road, Shanghai 200030, People’s Republic of China}

\author[0000-0002-8428-8050]{Aaron T. Lee}
\affiliation{Department of Physics and Astronomy, Saint Mary's College of California, 1928 Saint Mary's Road, Moraga, CA 94575, USA}

\author[0000-0003-1252-9916]{Stella S. R. Offner}
\affiliation{Department of Astronomy, University of Texas Austin, Austin, TX 78712, USA}

\author{James di Francesco }
\affiliation{NRC Herzberg Astronomy and Astrophysics,
5071 West Saanich Road,
Victoria, BC, V9E 2E7, Canada}
\affiliation{Department of Physics and Astronomy, University of Victoria,
3800 Finnerty Road, Elliot Building,
Victoria, BC, V8P 5C2, Canada}

\author[0000-0002-6773-459X]{Doug Johnstone}
\affiliation{NRC Herzberg Astronomy and Astrophysics,
5071 West Saanich Road,
Victoria, BC, V9E 2E7, Canada}
\affiliation{Department of Physics and Astronomy, University of Victoria,
3800 Finnerty Road, Elliot Building,
Victoria, BC, V8P 5C2, Canada}

\author[0000-0002-5809-4834]{Mika Juvela}
\affiliation{Department of Physics, P.O.Box 64, FI-00014,
University of Helsinki, Finland}

\author[0000-0002-6622-8396]{Paul F. Goldsmith}
\affiliation{Jet Propulsion Laboratory, California Institute of Technology, Pasadena CA 91109, USA.}

\author{Sheng-Li Qin}
\affiliation{Department of Astronomy, Yunnan University, Kunming 650091, People’s Republic of China}

\author{Xiaofeng Mai}
\affiliation{Shanghai Astronomical Observatory, Chinese Academy of Sciences, 80 Nandan Road, Shanghai 200030, People’s Republic of China}
\affiliation{School of Astronomy and Space Sciences, University of Chinese Academy of Sciences, No. 19A Yuquan Road, Beijing 100049, People’s Republic of China}

\author{Xun-chuan Liu}
\affiliation{Shanghai Astronomical Observatory, Chinese Academy of Sciences, 80 Nandan Road, Shanghai 200030, People’s Republic of China}

\author[0000-0002-7125-7685]{Patricio Sanhueza} %
\affiliation{National Astronomical Observatory of Japan, National Institutes of Natural Sciences, 2-21-1 Osawa, Mitaka, Tokyo 181-8588, Japan}
\affil{Department of Astronomical Science, SOKENDAI (The Graduate University for Advanced Studies), 2-21-1 Osawa, Mitaka, Tokyo 181-8588, Japan}

\author[0000-0001-5950-1932]{Feng-Wei Xu}
\affiliation{Department of Astronomy, School of Physics, Peking University, Beijing 100871, Peopleʼs Republic of China}
\affiliation{Kavli Institute for Astronomy and Astrophysics, Peking University, Haidian District, Beijing 100871, Peopleʼs Republic of China}

\author[0000-0002-8149-8546]{Ken'ichi Tatematsu}
\affil{Nobeyama Radio Observatory, National Astronomical Observatory of Japan,
National Institutes of Natural Sciences,
Nobeyama, Minamimaki, Minamisaku, Nagano 384-1305, Japan}
\affiliation{Department of Astronomical Science,
The Graduate University for Advanced Studies, SOKENDAI,
2-21-1 Osawa, Mitaka, Tokyo 181-8588, Japan}

\author[0000-0002-2338-4583]{Somnath Dutta}
\affiliation{Institute of Astronomy and Astrophysics, Academia Sinica, Roosevelt Rd, Taipei 10617, Taiwan (R.O.C)}

\author[0000-0002-9774-1846]{Huei-Ru Vivien Chen}
\affiliation{Institute of Astronomy and Department of Physics, National Tsing Hua University, Hsinchu, Taiwan
}

\author[0000-0003-1275-5251]{Shanghuo Li }
\affiliation{Max Planck Institute for Astronomy, K\"onigstuhl 17, 69117 Heidelberg, Germany}

\author[0000-0003-4546-2623]{Aiyuan Yang}
\affiliation{National Astronomical Observatories, Chinese Academy of Sciences, A20 Datun
Road, Chaoyang District, Beijing 100101, People’s Republic of China}

\author{Sheng-Yuan Liu}
\affiliation{Institute of Astronomy and Astrophysics, Academia Sinica, Roosevelt Rd, Taipei 10617, Taiwan (R.O.C)}

\author{Chin-Fei Lee}
\affiliation{Institute of Astronomy and Astrophysics, Academia Sinica, Roosevelt Rd, Taipei 10617, Taiwan (R.O.C)}

\author{Naomi Hirano}
\affiliation{Institute of Astronomy and Astrophysics, Academia Sinica, Roosevelt Rd, Taipei 10617, Taiwan (R.O.C)}

\author{Chang Won Lee}
\affiliation{Korea Astronomy and Space Science Institute, 776 Daedeokdae-ro, Yuseong-gu, Daejeon 34055, Republic of Korea}

\author[0000-0002-4393-3463]{Dipen Sahu}
\affiliation{Physical Research laboratory, Navrangpura, Ahmedabad, Gujarat 380009, India}
\affiliation{Academia Sinica Institute of Astronomy and Astrophysics, 11F of AS/NTU Astronomy-Mathematics Building, No.1, Sec. 4, Roosevelt Rd, Taipei 10617, Taiwan, R.O.C.}

\author[0000-0001-8385-9838]{Hsien Shang}
\affiliation{Institute of Astronomy and Astrophysics, Academia Sinica,  Taipei 10617, Taiwan}

\author[0000-0002-1369-1563]{Shih-Ying Hsu}
\affiliation{National Taiwan University (NTU), No. 1, Section 4, Roosevelt Rd, Taipei 10617, Taiwan (R.O.C.);seansyhsu@gmail.com}
\affiliation{Institute of Astronomy and Astrophysics, Academia Sinica, Roosevelt Rd, Taipei 10617, Taiwan (R.O.C)}

\author{Leonardo Bronfman}
\affiliation{Departamento de Astronomía, Universidad de Chile, Las Condes, 7591245 Santiago, Chile}

\author[0000-0003-4022-4132]{Woojin Kwon}
\affiliation{Department of Earth Science Education, Seoul National University, 1 Gwanak-ro, Gwanak-gu, Seoul 08826, Republic of Korea}
\affiliation{SNU Astronomy Research Center, Seoul National University, 1 Gwanak-ro, Gwanak-gu, Seoul 08826, Republic of Korea}

\author[0000-0002-6529-202X]{M. G. Rawlings}
\affiliation{Gemini Observatory/NSF’s NOIRLab, 670 N. A’ohoku Place, Hilo, Hawaii, 96720, USA}

\author{David Eden}  
\affiliation{Armagh Observatory and Planetarium, College Hill, Armagh, BT61 9DB, UK}

\author[0000-0003-2619-9305]{Xing Lu}
\affiliation{Shanghai Astronomical Observatory, Chinese Academy of Sciences, 80 Nandan Road, Shanghai 200030, People’s Republic of China}

\author{Qi-lao Gu}
\affiliation{Shanghai Astronomical Observatory, Chinese Academy of Sciences, 80 Nandan Road, Shanghai 200030, People’s Republic of China}

\author[0000-0003-4659-1742]{Zhiyuan Ren}
\affiliation{National Astronomical Observatories, Chinese Academy of Sciences, Datun Rd. A20, Beijing, People's Republic of China}

\author{D Ward-Thompson}
\affiliation{Jeremiah Horrocks Institute, UCLAN, Preston, PR12HE, UK}

\author{Zhi-Qiang Shen}
\affiliation{Shanghai Astronomical Observatory, Chinese Academy of Sciences, 80 Nandan Road, Shanghai 200030, People’s Republic of China}


\begin{abstract}
One of the most poorly understood aspects of low-mass star formation is how multiple-star systems are formed. 
Here we present the results of Atacama Large Millimeter/submillimeter Array (ALMA) Band-6 observations towards a forming quadruple protostellar system, G206.93-16.61E2, in the Orion B molecular cloud. ALMA 1.3\,mm continuum emission reveals four compact objects, of which two are Class I young stellar objects (YSOs), and the other two are likely in prestellar phase. The 1.3 mm continuum emission also shows three asymmetric ribbon-like structures that are connected to the four objects, with lengths ranging from $\sim$500 au to $\sim$2200 au. By comparing our data with magneto-hydrodynamic (MHD) simulations, we suggest that these ribbons trace accretion flows and also function as gas bridges connecting the member protostars.
Additionally, ALMA CO \textit{J}=2-1 line emission reveals a complicated molecular outflow associated with G206.93-16.61E2 with arc-like structures suggestive of an outflow cavity viewed pole-on.

\end{abstract}

\section{Introduction}\label{sec:intro}

Approximately half of the stars in the Galaxy reside in systems with two or more stars, with the multiplicity fraction for low-mass M stars being about one third \citep{2013Duchene, Offner_2022}. Near-infrared and sub-millimeter continuum observations towards star-forming regions in nearby molecular clouds have revealed that $\sim$28$\%$ to $\sim$64$\%$ of low-mass proto-stars are found in multiple systems, with typical separations of 100-8900 au between members \citep{Offner_2022}. Most of these young stellar systems, however, are binary systems, while higher-order systems are rare \citep{2013Chen, Kounkel2016, 2016Tobin,2022Tobin,2022Luo}.
Moreover, it is hard to determine whether higher-order systems are gravitationally bound based on continuum observations alone. \citet{Pineda2015} found a forming quadruple star system in the Barnard 5 dense core using NH$_3$ observations taken by the Very Large Array (VLA). The authors argued that Barnard 5 hosts a gravitationally bound system containing one protostar and three prestellar condensations with separations of a few thousand au. 
A dynamic analysis suggested that two wide-separated members in the systems will likely become unbound after formation.

Forming multiple systems are often observed to contain extended dynamic structures, such as gas streamers, and gas bridges connecting the members. The gas distribution and associated protostellar outflows in such systems are usually very complicated and thus difficult to interpret.
For example, the condensations in the Barnard 5 region are embedded in dense gas streamers \citep{Pineda2015}. Similar gas streamers have been witnessed in recent submillimeter observations toward other star-forming regions. In general, these gas streamers are asymmetric, with lengths of several to thousands of au \citep{2017Takawuwa,2020Rosotti,2020Keppler}.
\citet{2020Pineda} found a 10,500 au gas streamer in the Per-emb-2 dense core, which appears to be delivering material from the outer dense core to the inner disk. Computationally, a number of magneto-hydrodynamic (MHD) simulations have reproduced asymmetric streamers, e.g, gas bridges, which can be attributed to dynamical interaction between protostar members. For example, simulated flybys \citep{2010Kratter,2011Clark,201Kuffmeier, 2022Dong} produce bridges, though these bridge structures are caused by interactions between the forming stars, not the flyby per se.
Finally, observations of multiple protostellar systems have shown non-collimated, asymmetric, and complex outflow structures \citep{2021Oya,2015Kwon}.  
Hence, analyzing and understanding the dynamic mechanism of outflows and gas streamers is essential to unveiling the mystery of higher-order star system formation.

G206.93-16.61E2 is a protostellar core \citep{Yi2018} close to the reflection nebula NGC\,2023 in the Orion B molecular cloud (at a distance of 407\,$\pm$4\,pc, \cite{Kounkel2018}). The core is associated with the infrared source 2MASS J05413704-0217178, also known as HOPS-298 \citep{Megeath2012,2016Furlan}.
Recent ALMA Band-7 observations have resolved HOPS-298 into two Class I protostars (A, B) at a resolution of 0.1$^{\prime\prime}$ \citep{2020Tobin}. Besides these two protostars, the 1.3 mm continuum emission obtained by the ALMA Survey of Orion Planck Galactic Cold Clumps (ALMASOP) project further revealed two additional gas condensations within this core \citep{Dutta2020,2022Luo}. 

In this paper, we present spatially resolved observations of 1.3\,mm continuum, CO \textit{J}=2-1, C$^{18}$O (\textit{J}=2-1), SiO (\textit{J}=5-4), and H$_2$CO (\textit{J}=3-2) line emission toward G206.93-16.61E2. 
Our observations capture the presence of apparent streamer-like structures in the dust continuum emission.
In addition, we detect an outflow with a highly intricate dynamic behavior that has not been previously reported. The paper is organized as follows: In Section \ref{sec:obs} we describe the observations and data reduction. In Section \ref{sec:res} we present the observational results. We then discuss the origin of the quadruple system in Section \ref{sec:dis}. Finally, we summarize the conclusions in Section \ref{sec:sum}. 

\section{Observations} \label{sec:obs}

G206.93-16.61E2 was observed with ALMA at Band 6 from October 2018 to January 2019 as a part of the ALMASOP project (project ID: 2018.1.00302.S; PI: Tie Liu). The observations were conducted with the 12-m Array in its C43-5 (TM1) and C43-2 (TM2) configurations, and with the Atacama Compact Array (ACA). The receivers were set up to cover four individual spectral windows (centered at 216.6, 218.9, 231.0, and 233.0 GHz), each with a bandwidth of 1.875 GHz and a velocity resolution of 1.4 km s$^{-1}$. Molecular lines, including CO (\textit{J}=2-1), C$^{18}$O (\textit{J}=2-1), H$_2$CO \textit{J}=3-2, N$_2$D$^{+}$ (\textit{J}=3-2), DCO$^+$ (\textit{J}=3-2), DCN (\textit{J}=3-2) and SiO (\textit{J}=5-4), were simultaneously observed. These line-emission channels were removed from the ensemble of data used to make a continuum image. In this paper, we present the results of the 1.3\,mm continuum, CO (\textit{J}=2-1), C$^{18}$O (\textit{J}=2-1), SiO (\textit{J}=5-4), and H$_2$CO (\textit{J}=3-2) line emission data.

We performed the calibration and data imaging using the Common Astronomy Software Application (CASA) Version 5.4 \citep{McMullin2007}. 
The calibrated data were imaged with the TCLEAN task by combining data from all three configurations (TM1, TM2, and ACA) with a robust Briggs parameter of 0.5.
More details of the data reduction are described by \citet{Dutta2020}. The largest angular recoverable scale is $\sim$25$^{\prime\prime}$ and 
the field of view (FOV) of the final images is about 40$^{\prime\prime}$. The synthesized beam is 0.38$^{\prime\prime}$ $\times$ 0.33$^{\prime\prime}$ (152 au $\times$ 132au) and the sensitivity is $\sim$0.15 mJy\,beam$^{-1}$ in 1.3\,mm dust continuum emission. The final synthesized beam of $^{12}$CO J=2-1 is 0.36$^{\prime\prime}$ $\times$ 0.31$^{\prime\prime}$ (144 au $\times$ 124 au) based on the same setting, and the observations have an rms noise of 3 mJy\,beam$^{-1}$ per channel. 

\begin{figure*}
    \centering
    \includegraphics[width=18cm]{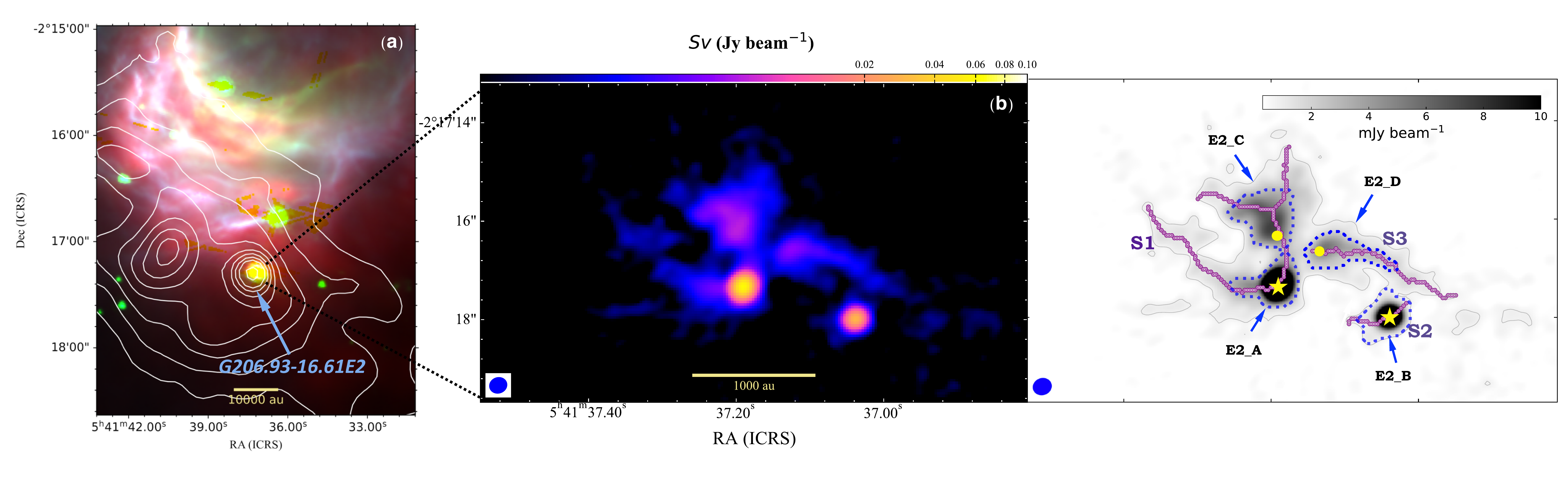}
    \caption{Dense core G206.93-16.61E2: (a) JCMT SCUBA-2 850\,$\mu$m in contours overlaid on RGB image of Herschel 100 $\mu$m (red) and Spitzer 4.5 (green) and 8\,$\mu$m (blue) data. The contours are given in white color from 4\,$\sigma_{0.85\,mm}$ to 28\,$\sigma_{0.85\,mm}$ with a step of 4\,$\sigma_{0.85\,mm}$ (1\,$\sigma_{0.85\,mm}$ = 25 mJy beam$^{-1}$).  (b) \textit{Left panel:} ALMA 1.3\,mm dust continuum image. \textit{Right panel:} the same image with one contour at the 5-$\sigma_{1.3\,mm}$ level in gray (1$\sigma_{1.3\,mm}$=0.15 mJy beam$^{-1}$). The blue dotted lines indicate the leaf structures, identified by Astrodendro, surrounding the sources E2$\_$A, E2$\_$B, E2$\_$C and E2$\_$D. The yellow symbols indicate the locations of the peak intensity in each structure. The yellow stars represent the two YSOs (E2$\_$A and E2$\_$B, also known as HOPS-298-A and B) and the yellow circles represent the two gas condensations (E2$\_$C and E2$\_$D). The continuum `ribbon' structures are indicated by purple lines. }
    \label{1.3mm}
\end{figure*}

\section{Results}\label{sec:res}

\subsection{ A forming quadruple stellar system} 

Figure \ref{1.3mm} shows the infrared and the dust continuum emission maps, for the G206.93-16.62E2 dense core obtained with various instruments. In Figure \ref{1.3mm} (b), ALMA 1.3\,mm observations spatially resolve four compact sources, which were previously reported by \citet{Dutta2020}. The two point sources G206.93-16.61E2$\_$A and G206.93-16.61E2$\_$B (hereafter E2$\_$A and E2$\_$B, respectively) correspond to the two Class I protostars, HOPS-298-A and HOP-298-B (projected separation $\sim$950 au), as also classified by \citet{2020Tobin}. The other two fainter and diffuse condensations, G206.93-16.61E2$\_$C (E2$\_$C) and G206.93-16.61E2$\_$D (E2$\_$D), were not detected by \citet{2020Tobin}.
Three of the sources, E2$\_$A, E2$\_$C, and E2$\_$D, are in close proximity to each other with a mean projected separation of about 450 au. The projected separation
between protostar E2$\_$B and the three other sources (E2$\_$A, E2$\_$C, and E2$\_$D)  are 950\,au, 1250\,au and 900\,au, respectively.

We utilize the Astrodendro Python package\footnote{https://dendrograms.readthedocs.io/en/stable/} to identify the boundaries of each source.
We adopt the min$\_$value = 3\,$\sigma_{1.3\,mm}$ (1\,$\sigma_{1.3\,mm}$ = 0.15 mJy beam$^{-1}$), min$\_$delta =  1\,$\sigma_{1.3\,mm}$, min$\_$npix = 18. The four members are defined by the smallest substructures, `leaf', in 1.3\,mm dust continuum emission, above the 5-$\sigma$ threshold found by Astrodendro. The resulting boundaries of these four sources are shown by the blue dotted shapes in the right panel of Figure \ref{1.3mm} (b). 
The four members are almost contiguous, and embedded within a common gas envelope.
Table \ref{members} lists the coordinates, flux densities, and deconvolved radii of these sources,  as extracted from Astrodendro. 

To determine whether the four individual sources are gravitationally bound,
we use the `leaf' boundaries derived above for each to derive the 1.3\,mm continuum-based gas mass $M_{gas}$, virial mass $M_{vir}$, and virial parameter $\alpha$ (see Appendix \ref{sec:cal}), which are also listed in Table \ref{members}. The gas masses ($M_{gas}$) of the four sources range from 0.1\,M${_\odot}$ to 0.39\,M${_\odot}$, assuming a dust temperature of 25 K for protostellar condensations (E2$\_$A and E2$\_$B) and 10 K for starless condensations (E2$\_$C and E2$\_$D). The gas masses of 
E2$\_$C ($\sim$0.39 M${_\odot}$) and E2$\_$D ($\sim$0.19 M${_\odot}$) are larger than their $M_{vir}$, indicating that the two condensations are likely gravitationally bound if they are as cold as 10\,K (See Appendix \ref{sec:cal}). Considering that no infrared or radio point sources have been detected toward E2$\_$C and E2$\_$D \citep{2020Tobin}, these two condensations are likely gas condensations in the prestellar phase.
Adopting a conservative 
star formation efficiency of 30\% estimated on core scales \citep{1998Andre,2017OffnerApJ...847..104O},  
we predict that E2$\_$C will have a final stellar mass slightly above the brown dwarf limit (80 Jupiter mass; $\sim$0.08\,M$_{\odot}$) and E2$\_$D might form a brown dwarf. However, both E2$\_$C and E2$\_$D may experience a rapid collapse within such a small ($\sim$200\,au) region,  resulting in a much higher in situ star formation efficiency. In addition, considering that E2$\_$C and E2$\_$D are embedded in continuum streamer-like structures (see Section \ref{streamer}), they might continue to accumulate gas from their surroundings. Therefore, both E2$\_$C and E2$\_$D may have potential to form a low-mass star.   

As discussed in Appendix~\ref{sec:lines}, C$^{18}$O \textit{J}=2-1 line emission is detected toward both E2$\_$C and E2$\_$D, but N$_2$D$^+$ \textit{J}=3-2 line emission as well as other lines targeted by the ALMASOP project (e.g., DCN, DCO$^{+}$) are not detected. This may indicate that the two starless gas condensations are warmer than 10 K. If we assume a higher temperature of 25 K for E2$\_$C and E2$\_$D, their gas masses decrease by a factor of $\sim$4, and virial masses increase by a factor of $\sim$1.5 (see Appendix~\ref{sec:cal}). In this case, the two gas condensations will not be gravitationally bound and may disperse in future. However, the observed C$^{18}$O \textit{J}=2-1 line emission does not resemble the continuum emission with compact structures, and is more likely related to the extended emission from the natal gas core at large scale. The non-detection of other lines is potentially due to the poor sensitivity and spectral resolution. Therefore, the two starless objects E2$\_$C and E2$\_$D may not be as warm as their protostellar counterparts in the system. Future enhanced sensitivity line observations from multiple transitions of known temperature probes (such as NH$_3$, HC$_3$N and CCS) are needed to better constrain the gas temperature of these objects. 

To summarize, G206.93-16.61E2 thus forms a quadruple stellar system consisting of two proto-stars and two candidate gravitationally bound gas condensations. This system is more compact, having much smaller separations between components, than the quadruple system in the Barnard 5 dense core \citep{Pineda2015}.

\begin{deluxetable*}{cccccccc}\footnotesize
\tablecaption{Properties of Members of G206.93-16.61E2\label{members}}
\tablehead{
\colhead{Source ID}&\colhead{R.A. (J2000)}& \colhead{Decl. (J2000)}  & \colhead{Flux density } & \colhead{Radius} & \colhead{$M_{gas}$}      & \colhead{$M_{vir}$}  &\colhead{$\alpha$} \\
\colhead{}               & \colhead{(hh:mm:ss)} & \colhead{(dd:mm:ss)}    & \colhead{(mJy)}            & \colhead{(au)}     & \colhead{(M$_{\odot}$)}  & \colhead{(M$_{\odot}$)} &\colhead{} 
}
\decimalcolnumbers
\startdata
G206.93-16.61E2$\_$A & 05:41:37.19 & −02:17:17.34 & 97 & 147 & 0.26 & 0.31 & 1.19 \\
G206.93-16.61E2$\_$B & 05:41:37.04 & −02:17:17.99 & 39 & 148 & 0.10 & 0.31 & 3.10 \\
G206.93-16.61E2$\_$C & 05:41:37.20 & −02:17:15.97 & 40 & 142 & 0.39 & 0.18 & 0.46\\
G206.93-16.61E2$\_$D & 05:41:37.15 & −02:17:16.52 & 19 & 141 & 0.19 & 0.19 & 1.00
\enddata
\tablecomments{Column (1)-(3): Names and coordinates of objects in G206.93-16.61E2, are taken from \citep{Dutta2020}. Column (4)-(5): Flux density and deconvolved radius of each object extracted from Astrodendro. Column (6)-(8): Gas mass, virial mass, and virial parameter are derived assuming a dust temperature of 25 K for protostellar objects and 10 K for starless objects (as discussed in Appendix \ref{sec:cal}, if the dust temperature of the starless objects is 25K then $M_{gas}$ decreases, $M_{vir}$ increases, and $\alpha$ increases). Gravitational collapse is expected when $\alpha<$ 2. }
\end{deluxetable*}

\subsection{Continuum `ribbons' around YSOs and condensations}\label{streamer}

Figure \ref{1.3mm} (b) clearly shows the presence of continuum streamer-like structures connecting the four compact objects, and two of the sources (E2$\_$A and E2$\_$D) are associated with very elongated features within the core. We used the FilFinder Python package\footnote{https://github.com/e-koch/FilFinder} to extract these elongated features from the mask created over 5$\sigma_{1.3\,mm}$ in 1.3\,mm dust continuum image. We adopt the branch$\_$thresh=\,450\,au, prune$\_$criteria=\,`length'. Three main continuum elongated structures (hereafter `ribbons'\footnote{We refer to the elongated structures seen in the dust continuum as `ribbons' rather than `streamer' in this work because we do not have complementary gas kinematic measurements associated with these structures.}), S1, S2, and S3 are identified, which are marked as purple lines in Figure \ref{1.3mm} (b). The three branches of S1 are considered to be substructures by Filfinder. The projected lengths of these continuum ribbons are 700\,au, 600\,au, and 2200\,au for S1, 500\,au for S2, and 1500\,au for S3.

The continuum ribbon S1 traverses both E2$\_$A and E2$\_$C, and bifurcates in the northern part of E2$\_$C. The continuum ribbon S2 crosses through E2$\_$B and may extend outward toward the western end of the continuum ribbon S3, while S3 connects E2$\_$D with the tail of its envelope. Since the length of the continuum ribbon S2 is comparable with the size of E2$\_$B and almost connects to S3, it may be a branch of S3 instead of a separate continuum ribbon. 
The continuum ribbons S1 and S3 converge into a hub region that forms E2$\_$A, E2$\_$C, and E2$\_$D, suggesting that these compact objects could continuously accumulate gas along these features. 

In previous studies, streamers have been often revealed by dense gas tracers, e.g., NH$_3$, HC$_3$N and CCS \citep{2019Alves,2019Yen,2020Ren,2020Pineda,2022Valdivia-Mena}, while a few additional similar structures are traced by continuum emission \citep{2016Perez,2021Sanhueza}. The three continuum ribbons in this system are clearly detected in 1.3\,mm dust emission. However, only a few molecular lines show weak emission at Band 6 in the ALMASOP observations (See Figure \ref{lines} in Appendix~\ref{sec:lines}), and none resemble the continuum emission. This may be due to both line excitation conditions and the sensitivity limit of the observations. Except for S2, the other two continuum ribbons, S1 and S3, extend to thousands of au in spatial scale and are likely formed by transporting material from the core to the protostars as funnels. These ribbons are likely only moderately heated by the embedded protostars and expected to be cold and may contain chemically fresh material \citep{2020Pineda}. 
\citet{2022Tatematsu}, found that the HCO$^+$ spectra exhibited inverse P Cyg-like absorption profiles toward the G206.93-16.61E2 dense core, suggesting that the core itself is collapsing. This also indicates that these continuum ribbons likely trace gas accretion. Future observations of low excitation lines, such as CCS and HC$_3$N, that trace chemically fresh material may help to reveal the gas kinematics to see whether the ribbon structures are formed via large-scale accretion flows funneling material down to disk scales or not.

\subsection{Molecular outflows}

\begin{figure*}
    \centering
    \includegraphics[width=18cm]{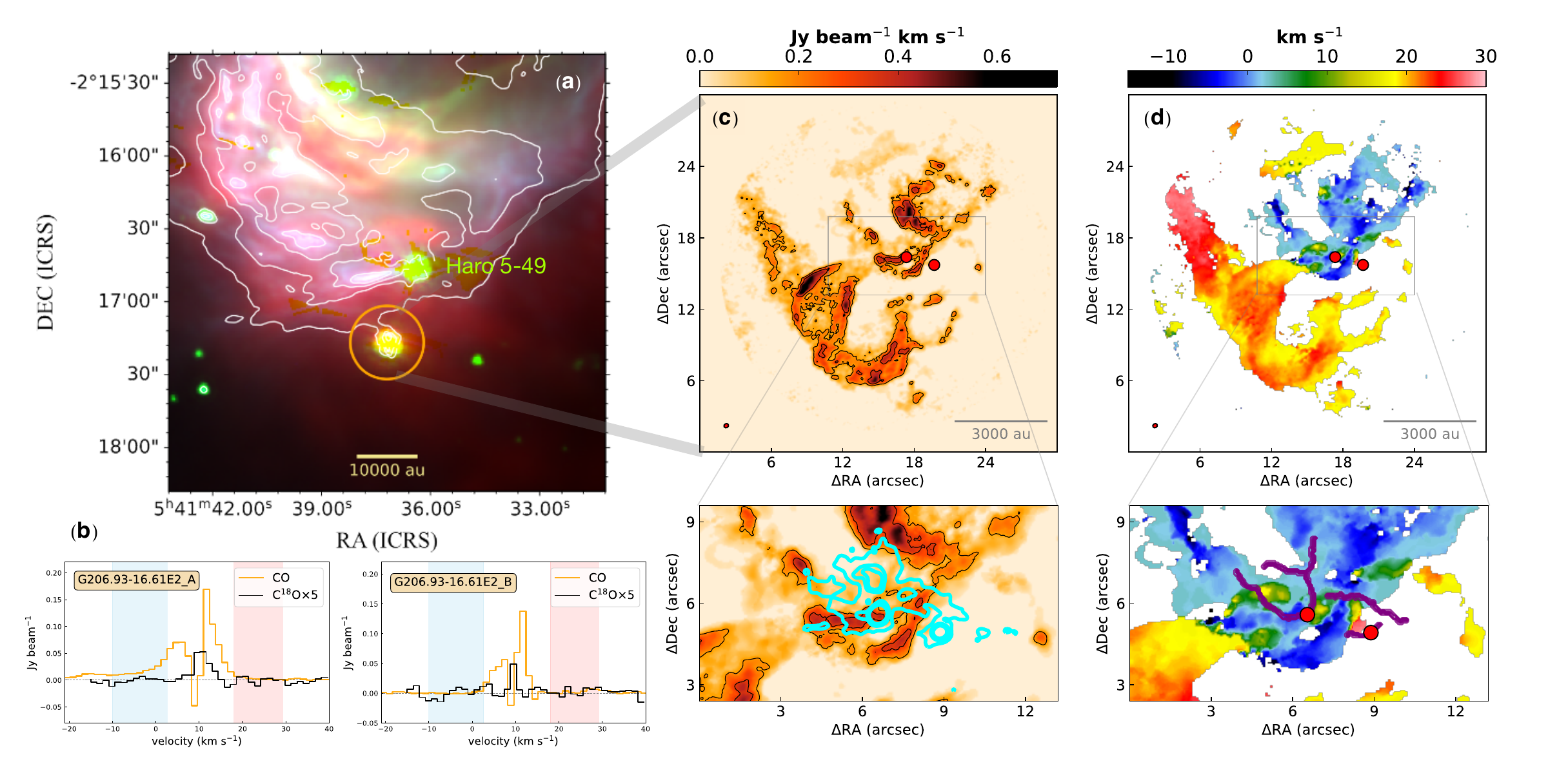}
    \caption{The CO outflow from G26.93-16.61E2: (a) The color image is the same as in Figure \ref{1.3mm}(a). The white contours show Spitzer 8\,$\mu$m emission. The orange circle shows the `field of view' (FOV) of the ALMA image; (b) C$^{18}$O and $^{12}$CO \textit{J}=2-1 spectra at the positions of G206.93-16.61E2$\_$A and G206.93-16.61E2$\_$B. The blue and red shaded areas define the blue- and red-shifted high velocity emission of $^{12}$CO line emission; (c) Moment-0 map of all high velocity emission combined ([-10 to 2.6] km s$^{-1}$ for blue and [18 to 29.2] km s$^{-1}$ for red) defined in panel (b), with contours of 10\,$\sigma_{out}$ and 25\,$\sigma_{out}$ (1\,$\sigma_{out}$=\,33 mJy beam$^{-1}$). The zoom-in picture shows the same image with dust continuum contours in cyan at 3\,$\sigma_{1.3mm}$ , 10\,$\sigma_{1.3mm}$ and 20\,$\sigma_{1.3mm}$; (d) Moment 1 map of high velocity emission. The red circles represent the protostar E2$\_$A and E2$\_$B. The structures of the continuum ribbons identified by Filfinder are shown in purple lines.
    }
    \label{outflow}
\end{figure*}

Recent studies have indicated that the outflows of multiple systems can have a great impact on the evolution of their member protostars  \citep{2022Jorgensen,2023Harada}.
In G206.93-16.61E2, the high-velocity $^{12}$CO \textit{J}=2-1 line emission reveals an outflow structure spanning 24$^{\prime\prime}$ ($\sim$9600 au), and consisting of two asymmetric, intricate arc-like structures around the protostellar system, as shown in Figure \ref{outflow}. The systemic velocity of G206.93-16.61E2 is 9.8 km\,s$^{-1}$ \citep{Kim2020}. To investigate the $^{12}$CO \textit{J}=2-1 and C$^{18}$O \textit{J}=2-1 line emission, we define the velocity ranges of [-10, 2.6] km s$^{-1}$ and [18, 29.2] km s$^{-1}$ for the blue- and redshifted high-velocity components of the $^{12}$CO \textit{J}=2-1 line wings (see Figure \ref{outflow}\,(b)). Figure \ref{outflow} (c) and (d) presents the moment maps for the high-velocity emission of the $^{12}$CO J=2-1 line: integrated intensity (Moment 0) and intensity-weighted velocity (Moment 1).

We calculate outflow parameters such as mass ($M_{\rm out}$), momentum ($P_{\rm out}$), energy ($E_{\rm out}$), projected length ($\lambda_{\rm out}$), dynamic timescale ($t_{\rm dyn}$), and mass outflow rate ($\dot{M}_{\rm out}$) from the $^{12}$CO \textit{J}=2-1 line wings.
The resulting values are listed in Table \ref{outflowtable}. The formulae used and more details for these calculations are described in Appendix \ref{sec:cal_outflow}. The total outflow mass and mass loss rate are 3.1 $\times10^{-3}$ M$_{\odot}$ and 2.5 $\times10^{-6}$ M$_{\odot}$~yr$^{-1}$, respectively. These results are similar to outflow parameters detected in the Orion A cloud \citep{2023Harada}. Note, however, that we adopt the maximum projected length of each lobe as the outflow length ($\lambda_ {\rm out}$). The actual lengths are larger than the projected value, and thus resulting in a mass loss rate that is an upper limit.

As shown in Figure \ref{outflow} (c), the outflow lobes do not exhibit a collimated bipolar outflow pattern. Instead, they appear to have divergent trajectories and arc-like structures. The outflow lobes are highly fragmented with strip and knot-like sub-structures. The overall morphology in G206.93-16.61E2 is reminiscent of the crescent-shaped structure of the pole-on outflow surrounding the protostar DK Cha \citep{2023Harada}. Given this, the complicated arc-like outflow structures in G206.93-16.61E2 may also indicate a nearly pole-on the outflow cavity. The pole-on direction of the outflow is also evidenced by the orientation of the disk. Both E2$\_$A and E2$\_$B show nearly circular-symmetric shapes in higher resolution ($\sim$40 au) continuum observations at 870 $\mu m$ \citep{2020Tobin}, indicating that their disks are likely viewed face-on.

Although this source (G26.93-16.61E2) is located at the edge of the reflection nebula NGC2023 (see Figure~\ref{outflow}(a)), the red and blue-lobe of the outflow do not appear to follow the shell of the nebula, indicating that the outflow is very unlikely induced by the expansion of the nebula. As shown in Figure~\ref{outflow}(c) and Figure~\ref{outflow}(d), the two detected YSOs (E2$\_$A and E2$\_$B) are located close to the center of the outflow lobes, and are thus likely candidate sources for driving the outflow. In particular, the protostar E2$\_$A is much closer to the outflow origin, and is likely the main driving source. The protostar E2$\_$B may be associated with partial weak outflow emission, although it is hard be confirmed based on the current data. The complicated structure of the outflow may be produced by the dynamical evolution of the quadruple system or turbulent motions during its accretion phase \citep{Offner2016}, which could be tested in future state-of-the-art numerical simulations.


\begin{deluxetable*}{rcccccccc}\footnotesize
\tablecaption{Outflow parameters of G206.93-16.61E2\label{outflowtable}}
\tablehead{
\colhead{Lobe}&
\colhead{V$_{lsr}$} & 
\colhead{$\Delta$ v } &
\colhead{M$_{out}$} & 
\colhead{P$_{out}$}& 
\colhead{E$_{out}$}&
\colhead{$\lambda$$_{out}$} &
\colhead{t$_{dyn}$} &  
\colhead{$\dot{M}$$_{out}$} 
\\ 
\colhead{}&
\colhead{(km s $^{-1}$)} & 
\colhead{(km s $^{-1}$)} & 
\colhead{(10$^{-2}$M$_\odot$)} & 
\colhead{(10$^{-2}$M$_\odot$km s $^{-1}$)}&
\colhead{(10$^{43}$erg)} &
\colhead{(10$^{-1}$pc)} &
\colhead{(10$^4$yr)} &
\colhead{(10$^{-5}$M$_\odot$yr$^{-1}$)} 
}
\decimalcolnumbers
\startdata
red  & 9.8 & [18,29.2] & 0.18 & 2.16 & 0.14 & 0.36 & 1.37 & 0.13 \\
blue & 9.8 & [-10,2.6] & 0.13 & 1.56 & 0.10 & 0.32 & 1.07 & 0.12
\enddata
\end{deluxetable*}

\section{Discussion}\label{sec:dis}

\begin{figure*}
    \centering
    \includegraphics[width=15cm]{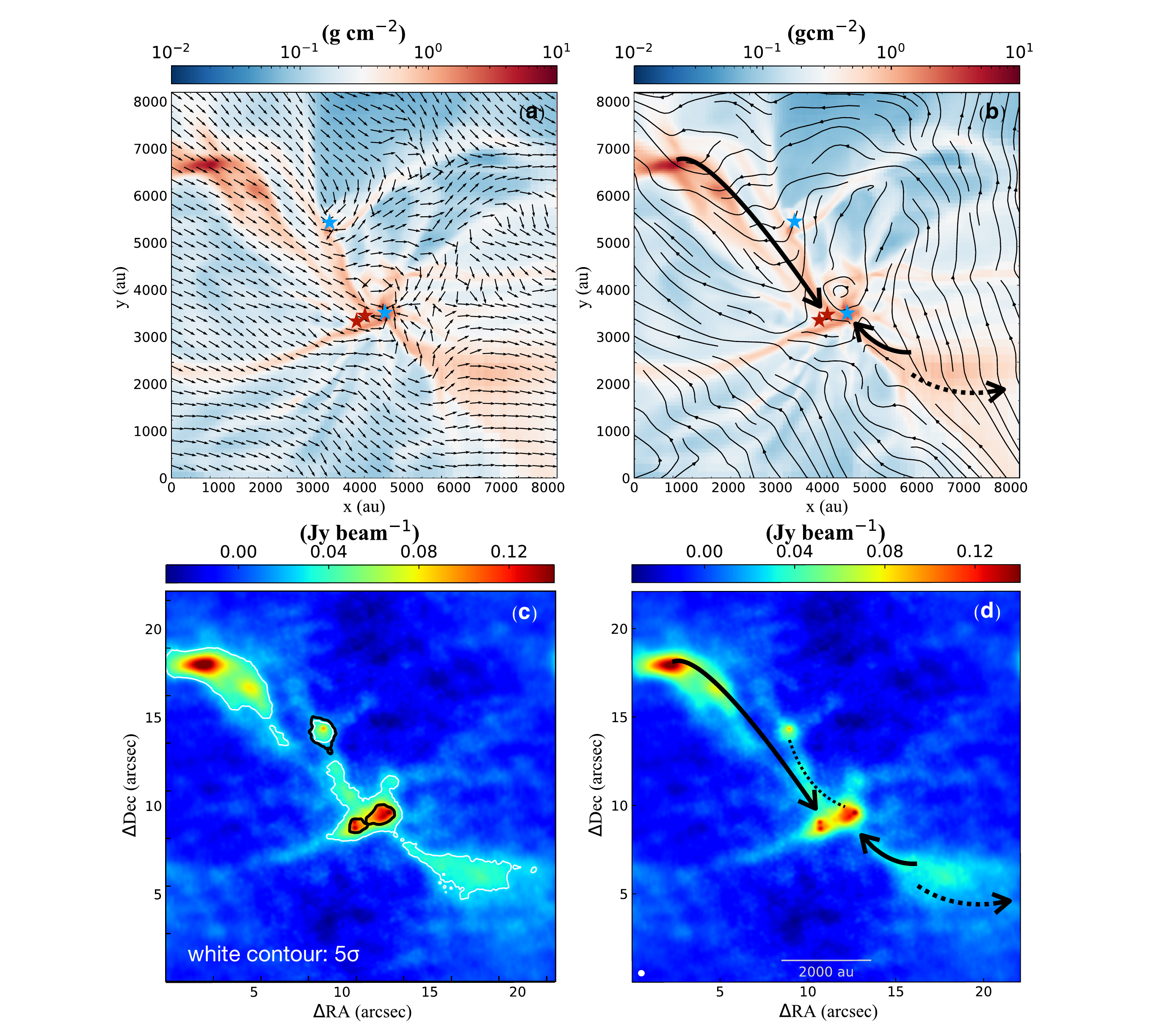}
    \caption{ Image of a quadruple star system simulated in three-dimensional and gravo-magnetohydrodynamic  \citep{2019Lee}. \textbf{(a)} H$_2$ column density map with the density-weighted projected velocity vectors (black arrows). The star symbols represent the location of the sink particles. Stars with the same colors were previously in a binary system; \textbf{(b)}  The color image is the same as in panel \textbf{(a)}. The black streamlines represent the density-weighted projected magnetic field. The black arrows indicate the direction of gas motion along the two major gas streamers; \textbf{(c)} Synthetic observation of 1.3\,mm dust continuum emission. The contour at 5\,$\sigma$ in white, and the black contour represents the `leaf' structure identified by Astrodendro (1\,$\sigma$ = 5\,mJy beam$^{-1}$); \textbf{(d)} The color image is the same as in panel \textbf{(c)}. The black solid lines with arrows represent the gas accretion flows along the streamers. The black dashed line marks a gas bridge connecting two protostars. }
    \label{lee}
\end{figure*}

Previous studies suggest that dynamical interactions among member stars can occur within star clusters/associations, potentially leading to the capture of members into bound systems \citep{2009Howe}.
Such events are seen in recent numerical simulations, and these results widely show the presence of gas-bridge streamers between protostars,which can be an evidence for mutual  evidence for mutual interacting processes between the stellar systems \citep{201Kuffmeier,2019Lee}. 

\citet{2019Lee} simulated several turbulent star-forming clouds with different magnetic field strengths to study how multiple systems form and evolve. 
Their simulations form a number of higher-order multiple systems, including a gravitationally bound quadruple that appears to be a good analog to G206.93-16.61E2. This system forms in the strong magnetic field cloud ($B_{\rm rms}\sim 32\, \mu$G). It has gas streamers and a hub-filament morphology as shown in Figure \ref{lee}(a) and (b). The lengths of the simulated gas streamer structures, 400 - 2500 au, are comparable to those in G206.93-16.61E2.
In their simulation, two proto-binary systems form via turbulent fragmentation and then merge to create a quadruple system, with the resulting streamers being a product of the interaction. Figure \ref{lee}(a) shows that gas material is mainly flowing along the upper-left streamers towards the protostars. The lower-right streamers are likely to be sheared with the close-in parts going inward and the rest moving outward. The gas flows along the major streamers are further illustrated with arrows in Figure \ref{lee}(b).

Figure \ref{lee} (c) and (d) display the synthetic 1.3\,mm continuum observation of the quadruple system in the simulation \citep{2019Lee}. To compare with our observations, we simulated 1.3 mm continuum emission assuming a dust temperature of 10 K and the optically thin continuum emission  (more details in Appendix \ref{sec:sim_data}) using the CASA \textit{simalma} task. 
We adopted the same antenna configurations (C43-2, C43-5, and ACA) and added thermal noise as used for the actual observations. Additionally, we added four point sources (0.1 Jy) representing protostars at the location of sink particles in \citet{2019Lee}. Two of the point sources are unresolved in the synthetic observation. The masses of the point sources identified in synthetic observations are 3.16, 1.37, and 1.47 \,M$_\odot$, respectively, which we note are massive than those in G206.93-16.61E2. The `leaf' structures identified by Astrodendro in the synthetic observations are shown in black contours in Figure \ref{lee} (c). These elongated structures surrounding point sources as also seen in G206.93-16.61E2 (see Figure~\ref{1.3mm}b) are likely shaped by the streamers around them.  As seen in Figure \ref{lee}(c) and (d), the synthetic observation at 1.3\, mm continuum clearly shows synthetic continuum-observed streamers of over 5\,$\sigma$ level around the protostars in dust emission. 
The continuum streamer marked by the black-dashed line in Figure \ref{lee}(d) is a gas-bridge connecting two protostars. The other continuum streamers marked by the black-solid arrows extend over thousands of au, tracing gas accretion flows that in some cases are transporting material from the surroundings to the protostars. 

Gas accretion along the upper-left streamer is also evidenced by pinched magnetic field as shown in Figure \ref{lee}(b). The magnetic field has been twisted along the accretion direction as marked by the black arrow. Close to the densest part of the streamer, the magnetic field is compressed and is roughly perpendicular to its spine. The accretion significantly increases the perpendicular component of the magnetic field at the streamer, forming a “U”-shaped field geometry. Such pinched magnetic field caused by gas accretion has been witnessed in both observations \citep{Liu2018c, Pillai2020} and other simulations works \citep{Gomez2018}. On the other hand, the magnetic field in the very central region is greatly distorted by gas accretion as well as interaction among member protostars. The magnetic field of the continuum ribbons in G206.93-16.61E2 could be investigated by high-sensitivity linear dust polarization observations with ALMA. With a continuum rms level of $\sim$10 $\mu$Jy~beam$^{-1}$, we expect to have a good detection of the linear polarization (S/N$>$3) even in its outer envelopes ($>0.75$
mJy~beam$^{-1}$ within the outer contours in the right panel of Figure \ref{1.3mm} ) that would have high fractional polarization \citep[5\%-11\% such as in][]{Cox2018,Maury2018}.

While this simulated system is not intended to replicate G206.93-16.61E2, their gas structures are remarkably similar. This supports the scenario that the continuum ribbons in G206.93-16.61E2 may act as accretion flows, and the continuum ribbon S2 may also function as bridge connecting member protostars, resulting from gravitational drag as the members migrate toward each other. This scenario can be further tested by future high-sensitivity and high-spectral resolution molecular line and dust polarization observations, which can reveal the details of the gas motions and pinched magnetic field of these continuum-observed ribbons.

\section{Conclusions}\label{sec:sum}

We have studied the young multiple stellar system G206.93-16.61E2 aimed to explore the formation mechanism of higher-order systems, using ALMA 1.3 mm dust continuum, and line emission data.  
The main results are summarized as follows:

1. G206.93-16.61E2 is forming a low-mass quadruple system. It contains two Class I protostellar sources and two candidate prestellar condensations. The four sources have separations smaller than 1000 au.

2. The 1.3 mm dust continuum emission reveals three distinct, asymmetric continuum ribbons with lengths ranging from 500-2200 au in G206.93-16.61E2. By comparing with MHD simulations, we suggest that these ribbons may trace gas accretion flows and also function as bridges connecting the members in this quadruple system. Future high-sensitivity and high-spectral resolution molecular line and dust polarization observations are needed to reveal the details of the gas motions and pinched magnetic field of these continuum-observed ribbons.

3. High-velocity $^{12}$CO line emission reveals an asymmetric, complicated outflow containing substructures such as strips and knots. Its arc-like structures likely indicate a pole-on outflow cavity.\\

\begin{acknowledgments}

This paper makes use of the following ALMA data: ADS/JAO.ALMA\#2018.1.00302.S. ALMA is a partnership of ESO (representing its member states), NSF (USA) and NINS (Japan), together with NRC (Canada), MOST and ASIAA (Taiwan), and KASI (Republic of Korea), in cooperation with the Republic of Chile. The Joint ALMA Observatory is operated by ESO, AUI/NRAO and NAOJ. This work has been supported by the National Key R$\&$D Program of China (No. 2022YFA1603101). Tie Liu acknowledges the supports by National Natural Science Foundation of China (NSFC) through grants No.12073061 and No.12122307, the international partnership program of Chinese Academy of Sciences through grant No.114231KYSB20200009, and Shanghai Pujiang Program 20PJ1415500. S.S.R.O. and A.T.L. are supported in part by NSF 1748571. The simulations of \citet{2019Lee} were done at the Massachusetts Green High Performance Computing Center (GHPCC LoneStar5) and the Texas Advance Computing Center (TACC Stampede2). 
D.J.\ is supported by NRC Canada and by an NSERC Discovery Grant.
This research was carried out in part at the Jet Propulsion Laboratory, which is operated by the California Institute of Technology under a contract with the National Aeronautics and Space Administration (80NM0018D0004).
X.L. acknowledges the supports by NSFC No. 12203086 and CPSF No. 2022M723278.
PS was partially supported by a Grant-in-Aid for Scientific Research (KAKENHI Number JP22H01271 and JP23H01221) of 
the Japan Society for the Promotion of Science (JSPS).
MJ acknowledges support from the the Academy of Finland grant No. 348342.
LB gratefully acknowledges support by the ANID BASAL project FB210003.
D.S. acknowledges the support from  Ramanujan Fellowship (SERB)  and PRL.
DWT wishes to thank the UK Science and Technology Facilities Council (STFC) for support under grant number ST/R000786/1.
\end{acknowledgments}

\appendix

\section{Masses and stability analysis}\label{sec:cal}

We derive the masses of the four G206.93-16.61E2 members assuming that 1.3 mm dust emission is optically thin, using the following formula:

\begin{equation}
    M_{gas}=\frac{S_{\nu}D^2}{\kappa_{\nu}B_{\nu}(T_{dust})},
\end{equation}
where, $S_{\nu}$ is the flux density from 1.3\,mm continuum emission from the ALMASOP observation and $D$ is the distance of 400 ${\rm pc}$. $\kappa_{\nu}$ is the dust opacity per unit mass column density at 1.3\,mm, and the parameters are adopted from \citet {Dutta2020}.
We assume that the dust temperature of the protostellar condensations E2$\_$A and E2$\_$B is $T_{\rm dust}$ =25\,K.For the condensations E2$\_$C and E2$\_$D, based on their unknown evolutionary status, we derived the gas mass with $T_{\rm dust}$ ranging from 10\,K to 25\,K. 

To determine the virial mass, we adopt the N$_2$H$^+$ line width of the core obtained from Nobeyama 45-m (NRO-45m) observations to derive the total velocity dispersion of N$_2$H$^+$, $\sigma_{v_{N_2H^+}}$ = $\frac{\Delta V_{N_2H^+}}{\sqrt{8ln2}}$ $\sim$ 0.24\,km s$^{-1}$, $\Delta V_{N_2H^+}$ is the full width at half maximum (FWHM) from \cite{Kim2020}. 
 Given that the N$_2$H$^+$ data have a much larger beam size of $\sim$19$^{\prime\prime}$ than that of ALMA, which covers a large portion of the dense core, the derived 3-d gas line width should be treated as an upper limit in the following virial analysis for the gas condensations detected by ALMA. Assuming a density profile $\rho$ $\propto$ $r^{-1.5}$, we calculate the virial masses of the gas condensations following \citet{1994AWilliams}:

\begin{equation}
     M_{vir} = \frac{5R\sigma^{2}_{v_{3D}}}{3\gamma G},
\end{equation}
where, $G$ is the gravitational constant, $\gamma$ = $\frac{4}{5}$ when $\rho$ $\propto$ $r^{-1.5}$, and $R$ is the deconvolved radius of each member provided by Astrodendro following the procedure by \citet{2023Xu}. $\sigma_{v_{3D}}$ is the 3-dimensional velocity dispersion for the H$_2$ gas, $\sigma_{v_{3D}}$ = $\sqrt{3(\sigma^2_{TH} + \sigma_{NT}^2})$. 
The thermal velocity dispersion of the H$_2$ gas is $\sigma_{TH}$ = $\sqrt{(\frac{k_b T_{k}}{m_H\mu})}$, where $k_b$ is the Boltzmann's constant, $T_{k}$ is the kinetic temperature, $m_H$ is the hydrogen mass, and $\mu$ is the molecular weight of the H$_2$. 
The non-thermal velocity dispersion, $\sigma_{NT}$, is derived from the full width at half maximum (FWHM) of the N$_2$H$^+$ line width, given by $\sigma_{NT}$ = $\sqrt{\sigma^2_{v_{N_2H^+}}-\sigma^2_{TH,N_2H^+}}$. 
The thermal velocity dispersion of N$_2$H$^+$ ($\sigma_{TH,N_2H^+}$) is derived by assuming a gas temperature $\sim$10K-25K. The velocity dispersion, gas masses and virial masses for the two starless gas condensations E2$\_$C and E2$\_$D are summarized in Table~\ref{starless}.

\begin{table}[]
\caption{Physical parameters of the two starless gas condensations}\label{starless}
\begin{tabular}{c|cccccc|cccccc}
\hline
\hline
{Source ID} & \multicolumn{6}{c|}{T$_{dust}$ = 10K}                                                                 & \multicolumn{6}{c}{T$_{dust}$ = 25K}                                                                  \\ \cline{2-13} 
                           & $\sigma_{TH}$ & $\sigma_{NT}$ & \multicolumn{1}{c|}{$\sigma_{3D}$} & $M_{vir}$ & \multicolumn{1}{c|}{$M_{gas}$} & $\alpha$ & $\sigma_{TH}$ & $\sigma_{NT}$ & \multicolumn{1}{c|}{$\sigma_{3D}$} & $M_{vir}$ & \multicolumn{1}{c|}{$M_{gas}$} & $\alpha$ \\
                           & \multicolumn{3}{c|}{(km s$^{-1}$)}                                 & \multicolumn{2}{c|}{(M$_{\odot}$)} & & \multicolumn{3}{c|}{(km s $^{-1}$)}                                & \multicolumn{2}{c|}{(M$_{\odot}$)} & \\ \hline
G206.93-16.61E2$\_$C       & 0.20          & 0.24          & \multicolumn{1}{c|}{0.54}          & 0.18      & \multicolumn{1}{c|}{0.39}      & 0.46     & 0.32          & 0.23          & \multicolumn{1}{c|}{0.68}          & 0.29      & \multicolumn{1}{c|}{0.11}      & 2.63     \\ \cline{1-1}
    G206.93-16.61E2$\_$D       & 0.20          & 0.24          & \multicolumn{1}{c|}{0.54}          & 0.19      & \multicolumn{1}{c|}{0.19}      & 1.00     & 0.32          & 0.23          & \multicolumn{1}{c|}{0.68}          & 0.30      & \multicolumn{1}{c|}{0.05}      & 6.00     \\ \hline
\end{tabular}
\end{table}

The virial parameter $\alpha$ of the each member is derived using the equation:
\begin{equation}
   \alpha = M_{vir} / M_{gas}.
\end{equation}
 A source is gravitationally bound if $\alpha<2$ \citep{2013Kauffmann}. For the condensations E2$\_$C and E2$\_$D, the gas mass $M_{gas}$ and the virial parameter $\alpha$ are calculated for two different dust temperature conditions. We list the derived values for $M_{gas}$ and $\alpha$ at 10K and 25K in Table \ref{starless}.

\begin{figure}
    \centering
    \includegraphics[width=8cm]{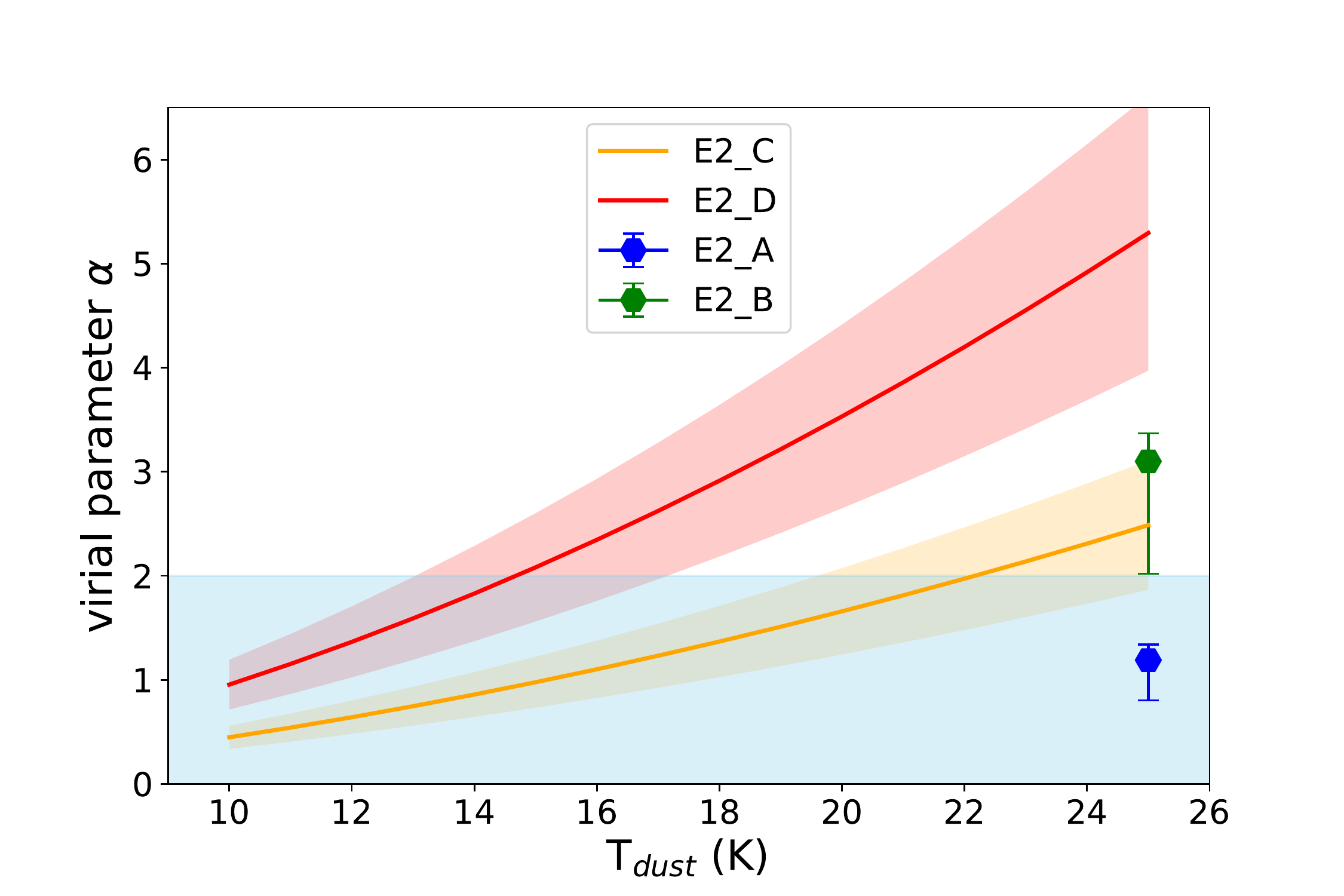}
    \caption{The dust temperature versus virial parameter for the four gas condensations in G206.93-16.61E2. The blue and green polygons with error bar denote the virial parameters of E2$\_$A and E2$\_$B at 25\,K, respectively. The orange and red lines with their hatched regions show the corresponding virial parameters and dust temperature T$_{dust}$ values ranging from 10\,K to 25\,K. The lines correspond the good model for describing infalling envelopes $\rho \propto \gamma^{1.5}$. The hatched regions and error bars correspond to the the density profile models from $\rho \propto \gamma^{0}$ - $\gamma^{2}$.
    The blue regions highlights the virial parameters $\alpha$ $\le$ 2 that indicate a gravitationally bound status. }
    \label{dust}
\end{figure}

Figure \ref{dust} shows the virial parameters as a function of dust temperature for the four gas condensations.
E$\_$A and E2$\_$B have been classified as Class I protostars. The virial parameter of E2$\_$A is below 2 while E2$\_$B is above 2. Since we do not take into account their stellar masses, the virial parameters of E2$\_$A and E2$\_$B should be overestimated.
For the two starless condensations E2$\_$C and E2$\_$D, their virial parameters increase as the dust temperature increases. Their virial parameters exceed the boundary of $\alpha=2$ at $\sim14$ K and $\sim22$ K, respectively.
The accurate temperature and corresponding virial parameters of E2$\_$C and E2$\_$D should be further constrained in future observations. 

\section{Molecular line emission}\label{sec:lines}

\begin{figure*}
    \centering
    \includegraphics[width=18cm]{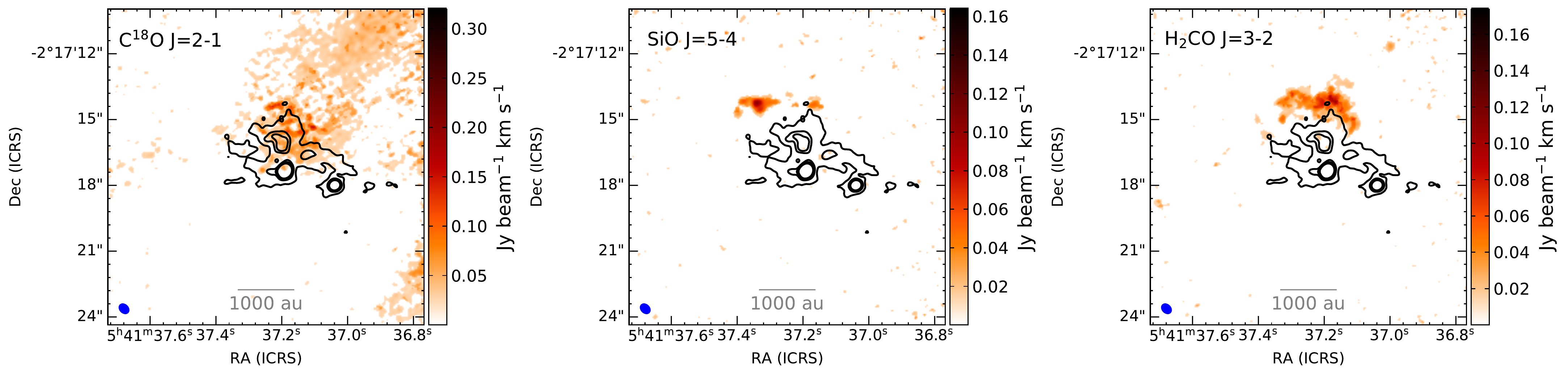}
    \caption{ Integrated intensity maps of C$^{18}$O, SiO and H$_{2}$CO emission, with dust continuum in black contours at 3\,$\sigma_{1.3\,mm}$, 10\,$\sigma_{1.3\,mm}$ and 20\,$\sigma_{1.3\,mm}$. }    
    \label{lines}
\end{figure*}

Figure \ref{lines} presents the integrated intensity maps for C$^{18}$O \textit{J}=2-1, SiO \textit{J}=5-4, and H$_2$CO \textit{J}=3-2. These emission maps do not have a direct correspondence to the continuum `ribbons'. The C$^{18}$O emission exhibits an extended feature and does not show any compact emission that is associated with the four gas condensations, indicating that its emission may originate from the large-scale natal gas core. The emission from both SiO and H$_2$CO is concentrated to the north of the continuum emission, potentially denoting a shell-like structure that may be induced by outflow shocks. 

\section{Physical properties of the outflow}\label{sec:cal_outflow}

To estimate the physical properties of the outflow we assume that high-velocity CO emission is optically thin and use the following equations to calculate the outflow parameters: outflow mass (\textit{M}$_{out}$), momentum (\textit{P}$_{out}$), energy(\textit{E}$_{out}$), dynamic timescale (\textit{t}$_{dyn}$), projected length ($\lambda_{out}$), and mass outflow rate (\textit{$\dot{M}$} $_{out}$) \citep {2009Qiu,2020Lis}. 

\begin{equation*}
M_{out}= 1.36\times10^{-6}exp\left(\frac{16.59}{T_{ex}}\right)\left(T_{ex}+0.92\right)D^{2}\int{\frac{\tau_{12}({1-e^{-\tau_{12}}})}{S_{\nu}}d\nu};
\end{equation*}

\begin{equation*}
P_{out}=\sum M_{out}(v)v;
\end{equation*}

\begin{equation*}
E_{out}=\frac{1}{2}\sum M_{out}(v)v^2;
\end{equation*}

\begin{equation*}
t_{dyn}=\frac{\lambda}{(v_{max(b)} + v_{max(r)})/2};
\end{equation*}

\begin{equation*}
\dot{M}_{out}=\frac{M_{out}}{t_{dyn}}.
\end{equation*}
%
%
%
Here, we assume excitation temperature $T_{ex}$ = 30\,K,  D represents the distance in kpc $\sim$ 0.40\textit{pc}, $M_{out}$ has a unit of M$_{\odot}$, $S_\nu$ is the flux density from the $^{12}$CO emission, and $v_{max}$ is the maximum velocity of each lobe.

\section{Flux density of the simulated data}\label{sec:sim_data}

The 1.3\,mm flux density is determined using the H$_2$ column density \citep{2007Kauffmann}, $N_{\rm H_2}$, assuming the dust to be optically thin, i.e,

\begin{equation*}
F_{\nu} = 2.02\times10^{-20}{\rm cm}^{-2}(e^{1.439(\lambda/{\rm mm})^{-1}(T/10K)^{-1}}-1)^{-1}(\frac{\lambda}{{\rm mm}})^{-3}(\frac{\kappa_\nu}{0.01{\rm cm}^2~{\rm g}^{-1}})(\frac{\theta_{\rm pixel}}{10~{\rm arcsec}})^{2}\times N_{H_2}.
\end{equation*}
Here, $F_{\nu}$ has a unit of Jy pixel$^{-1}$, $\lambda$ is the wavelength of the continuum in mm, $T_{\rm dust}$ is the dust temperature,  which we assume is 10\,K, $\kappa_{\nu}$ is the dust opacity from Equation (A1), and $\theta_{\rm pixel}$ is the angular pixel size.


%

\vspace{5mm}



\bibliography{output.bbl}{}
\bibliographystyle{aasjournal}






\end{document}